\newcommand{\modelname}{\textsf{KG-Diverse}\xspace}

\documentclass[sigconf]{acmart}
\usepackage{natbib}
\usepackage{amsmath,amsfonts}
\usepackage{algorithmic}
\usepackage{graphicx}
\usepackage{textcomp}
\usepackage{multirow,makecell}
\usepackage{enumitem}
\usepackage{xspace}
\usepackage[normalem]{ulem}
\useunder{\uline}{\ul}{}
\usepackage{color}
\usepackage{amsmath}
\usepackage{subcaption}
\usepackage{algorithm}

\AtBeginDocument{%
  \providecommand\BibTeX{{%
    \normalfont B\kern-0.5em{\scshape i\kern-0.25em b}\kern-0.8em\TeX}}}
\copyrightyear{2024}
\acmYear{2024}
\setcopyright{acmlicensed}\acmConference[WSDM '24]{Proceedings of the 17th ACM International Conference on Web Search and Data Mining}{March 4--8, 2024}{Merida, Mexico}
\acmBooktitle{Proceedings of the 17th ACM International Conference on Web Search and Data Mining (WSDM '24), March 4--8, 2024, Merida, Mexico}
\acmPrice{15.00}
\acmDOI{10.1145/3616855.3635803}
\acmISBN{979-8-4007-0371-3/24/03}

\begin{document}

\title{Knowledge Graph Context-Enhanced Diversified Recommendation}

\author{Xiaolong Liu}
\email{xliu262@uic.edu}
\affiliation{%
  \institution{University of Illinois Chicago}
  \city{Chicago}
  \country{USA}}
  
\author{Liangwei~Yang}
\email{lyang84@uic.edu}
\affiliation{%
  \institution{University of Illinois Chicago}
  \city{Chicago}
  \country{USA}}

\author{Zhiwei~Liu}
\affiliation{%
  \institution{Salesforce AI Research}
  \city{Palo Alto}
  \country{USA}
}
\email{zhiweiliu@salesforce.com}
  
\author{Mingdai~Yang}
\email{myang72@uic.edu}
\author{Chen~Wang}
\email{cwang266@uic.edu}
\affiliation{%
  \institution{University of Illinois Chicago}
  \city{Chicago}
  \country{USA}}

\author{Hao Peng}
\affiliation{%
   \institution{School of Cyber Science and Technology, Beihang University,}
   \country{Beijing, China}\\
   \institution{Yunnan Key Laboratory of Artificial Intelligence, Kunming University of Science and Technology,}
   \country{Kunming, China}}
\email{penghao@buaa.edu.cn}
\authornote{Corresponding author}

\author{Philip S.~Yu}
\affiliation{%
  \institution{University of Illinois Chicago}
  \city{Chicago}
  \country{USA}}
\email{psyu@uic.edu}

\renewcommand{\shortauthors}{Xiaolong Liu et al.}

\begin{abstract}
The field of Recommender Systems (RecSys) has been extensively studied to enhance accuracy by leveraging users' historical interactions. Nonetheless, this persistent pursuit of accuracy frequently engenders diminished diversity, culminating in the well-recognized "echo chamber" phenomenon. Diversified RecSys has emerged as a countermeasure, placing diversity on par with accuracy and garnering noteworthy attention from academic circles and industry practitioners.
This research explores the diversified RecSys within the intricate context of knowledge graphs (KG). These KGs act as repositories of interconnected information concerning entities and items, offering a propitious avenue to amplify recommendation diversity through the incorporation of insightful contextual information. Our contributions include introducing an innovative metric, Entity Coverage, and Relation Coverage, which effectively quantifies diversity within the KG domain. Additionally, we introduce the Diversified Embedding Learning (DEL) module, meticulously designed to formulate user representations that possess an innate awareness of diversity. In tandem with this, we introduce a novel technique named Conditional Alignment and Uniformity (CAU). It adeptly encodes KG item embeddings while preserving contextual integrity. Collectively, our contributions signify a substantial stride towards augmenting the panorama of recommendation diversity within the KG-informed RecSys paradigms.
We release the code at \textcolor{blue}{\url{https://github.com/Xiaolong-Liu-bdsc/KG-diverse}}.
\end{abstract}



\maketitle

\section{Introduction}

Recommender system (RecSys) has emerged as a vital solution for mitigating information overload in the contemporary landscape of extensive data~\cite{bigdata}. 
RecSys acquires the underlying preference by gaining insights from historical user-item interactions~\cite{liu2023group,yang2023unified}.
Subsequently, it offers personalized suggestions from a pool of potential items. Its pervasiveness is evident across multiple facets of our quotidian existence, encompassing domains such as news feeds~\cite{npa}, cinematic suggestions~\cite{gomez2015netflix}, and e-commerce recommendation~\cite{schafer2001commerce}. 

Predominantly driven by the imperative of accuracy, commercial entities~\cite{cheng2016wide,gomez2015netflix,ying2018graph,liu2022monolith, TA-HGAT,wang2023exploiting,wang2022contrastvae,wang2023conditional} have endeavored to construct intricate algorithms. 
These algorithms predict the item that most likely engages each user, predicated upon their historical interactions. 
The mechanisms~\cite{zhou2010solving} striving for a balance between accuracy and diversity have been developed to address this issue.
This deficiency in diversity results in the echo chamber or filter bubble phenomenon~\cite{echochamber}, where users are repeatedly exposed to content aligning with their preferences, limiting their exposure to diverse options.


The paradigm of Diversified Recommender Systems (Diversified RecSys) has emerged to counteract the aforementioned constraints by actively prioritizing diversity during the recommendation process~\cite{DGCN,DGRec}. 
These systems incite users to embark upon a broader exploration of choices. Consequently, users are enabled to encounter novel items and potentially discover items of interest that might have otherwise eluded their attention. Diversified RecSys has garnered escalating scholarly and industrial attention~\cite{DGCN,DGRec,cheng2017learning,chen2018fast}. Given the inherent conundrum between diversity and accuracy~\cite{zhou2010solving}, these systems are geared towards optimizing diversity while keeping the compromise on accuracy to a minimum, thus yielding an improved trade-off. Current methods gauge diversity through the lens of item categorization attributes, such as category coverage~\cite{DGCN,DGRec,chen2018fast}. This approach primarily operates at a coarse-grained categorical level, thereby exhibiting limitations in comprehensively assessing diversity. 
It can not adequately discriminate between recommending items from within the same category. 
For example, recommending both ("iPhone 14" and "Galaxy S23") and ("iPhone 14" and "iPhone 14 Pro") yields identical diversity outcomes due to their shared electronic category, which falls short of a comprehensive measure for diversification.


To cope with this, we delve into the diversified RecSys under the knowledge graphs (KG) framework. 
Extensive literature has been published on recommender systems that incorporate KG~\cite{KGAT, KGCL,KGIN,KACL}.
To diversify RecSys, a primary focal point of this investigation revolves around the nuanced integration of knowledge graphs. This entails leveraging the intricate details within the knowledge graph. 
To illustrate this, consider the previous example. In this context, a recommendation such as ("iPhone 14," "Galaxy S23") acquires heightened diversification by enlisting entities encompassing not only the specific products but also broader contextual elements, including nations ("United States," "Korea"), manufacturers ("Apple," "Samsung"), and operating systems ("iOS," "Android"), all drawn from the encompassing knowledge graph. As a result, users are endowed with an augmented sphere of exposure to a diverse spectrum of entities, thereby enriching their engagement through the recommended selections.


Effective enhancement of recommendation diversity through incorporating KG information presents several pivotal challenges that merit careful consideration. 
(1) Actively prioritizing diversity during the recommendation process with KG: 
The intrinsic nature of KG structures poses a distinctive predicament where KG entities do not directly encapsulate user attributes. This inherent disconnect consequently engenders a deficiency in accurately representing user embeddings derived from KG information. Consequently, proficiently depicting and diversifying user embeddings using KG data constitutes a significant challenge. 
(2) Striking a Balance between Accuracy and Diversity:
Higher accuracy often comes at the cost of reduced diversity.
For all diversified RecSys, the inherent challenge lies in attaining elevated diversity without compromising recommendation accuracy.
(3) Inadequate Characterization of Item KG Context Similarity: Present methodologies~\cite{KGAT,KGIN,wang2022metakrec} exhibit limitations in the robust encoding of item KG context similarity. This deficiency impairs the capacity to comprehensively discern item similarity from the vantage point of the KG. The apt assessment of item similarity within the KG milieu bears particular significance when striving to impart diversity to recommendations.

\begin{figure}[htbp]
    \centering
    \includegraphics[scale=0.38]{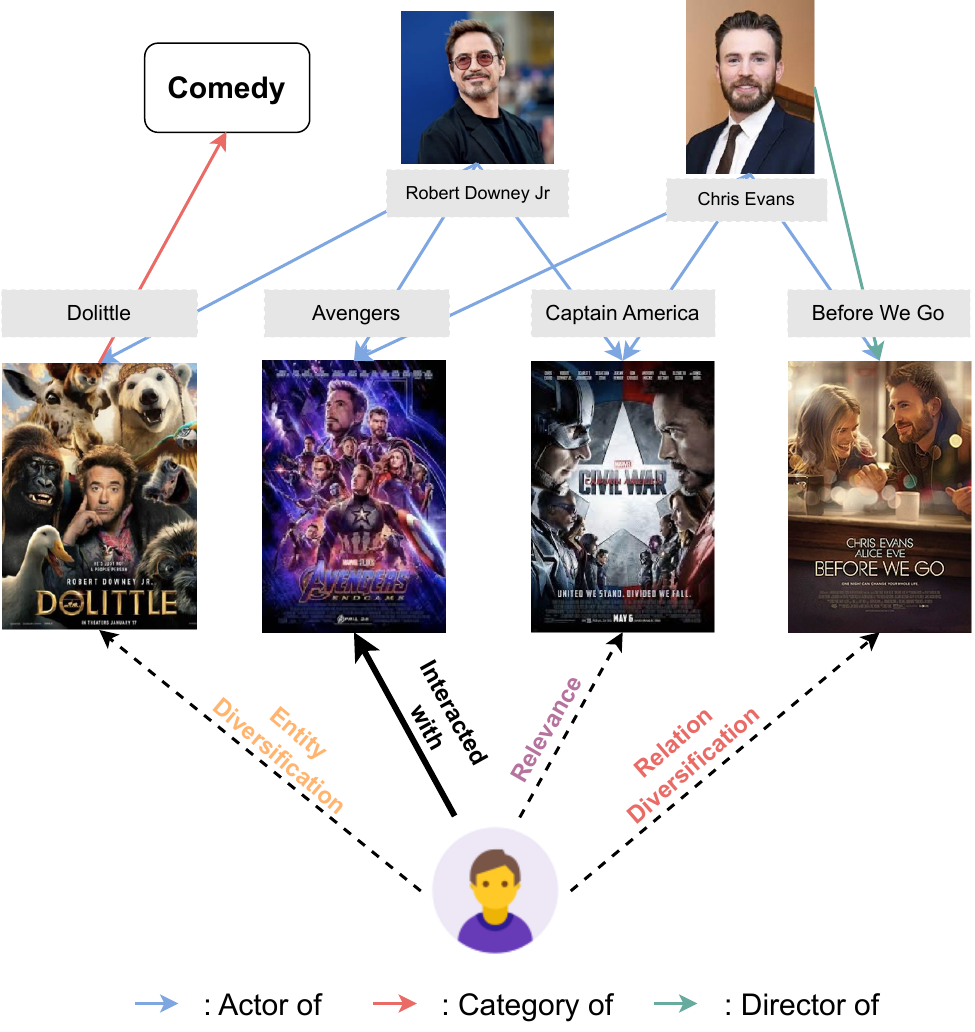}
    \caption{Diversified RecSys with knowledge graph.}
    \label{fig:mesh1}
\end{figure}

In this study, we introduce a novel framework, denoted as \modelname, aimed at addressing the aforementioned challenges. The cornerstone of our approach involves the formulation of two comprehensive metrics for gauging recommendation diversity within Knowledge Graphs (KGs): (1) Entity Coverage (EC) and (2) Relation Coverage (RC). These metrics assess the extent to which recommended items encapsulate a wide array of entities and relations within the KG context.
To illustrate the practical implications of our proposed metrics, consider the scenario depicted in Figure~\ref{fig:mesh1}, wherein a user's viewing history encompasses the film~\textit{The Avengers}. A conventional RecSys would potentially propose~\textit{Captain America 3} owing to the shared lead actors~\textit{Chris Evans} and~\textit{Robert Downey Jr.} with~\textit{The Avengers}, thus highlighting a thematic relevance. However, the introduction of diversified recommendations such as the film~\textit{Dolittle}, notable for its comedic genre and featuring actor~\textit{Robert Downey Jr.}, or the movie~\textit{Before We Go}, notable for actor~\textit{Chris Evans} assuming both acting and directorial roles, serves to introduce elements of entity and relation diversification, respectively. Such nuanced recommendations contribute to a more holistic and enriching user experience.
Consequently, we proceed to present an innovative module coined the Diversified Embedding Learning (DEL) module. This module is thoughtfully devised to engender personalized user representations imbued with an awareness of diversity, thereby fostering the augmentation of recommendation diversity without compromising recommendation accuracy.
The underpinning of effective KG embedding learning is fundamentally pivotal for diversification. In response, we proffer a novel strategy labeled "conditional alignment and uniformity." This strategy is meticulously designed to uphold the integrity of KG embeddings and preserve the intrinsic similarity between items that share common entities. This dual-pronged approach ensures both the quality of KG embeddings and the coherence of item similarity within the KG context.
Our contributions are summarized as follows:
\begin{itemize}[leftmargin=*]
    \item To the best of our knowledge, this paper firstly introduces the novel measurement of recommendation diversity in KG through the use of Entity Coverage and Relation Coverage metrics.
    \item We propose a simple yet effective Diversified Embedding Learning module to generate diversity-aware representations for users. 
    Additionally, we design a novel technique, conditional alignment and uniformity, to effectively encode item embeddings in KG.
    \item We evaluate the recommendation performance of our method on three public datasets. The extensive experimentation validates the effectiveness of our proposed model, showing the significance of each module in augmenting diversity while incurring only a negligible decrease in the accuracy of recommendations.
\end{itemize}

\section{Methodology}

\begin{figure*}[htbp]
    \centering
    \includegraphics[scale=0.396]{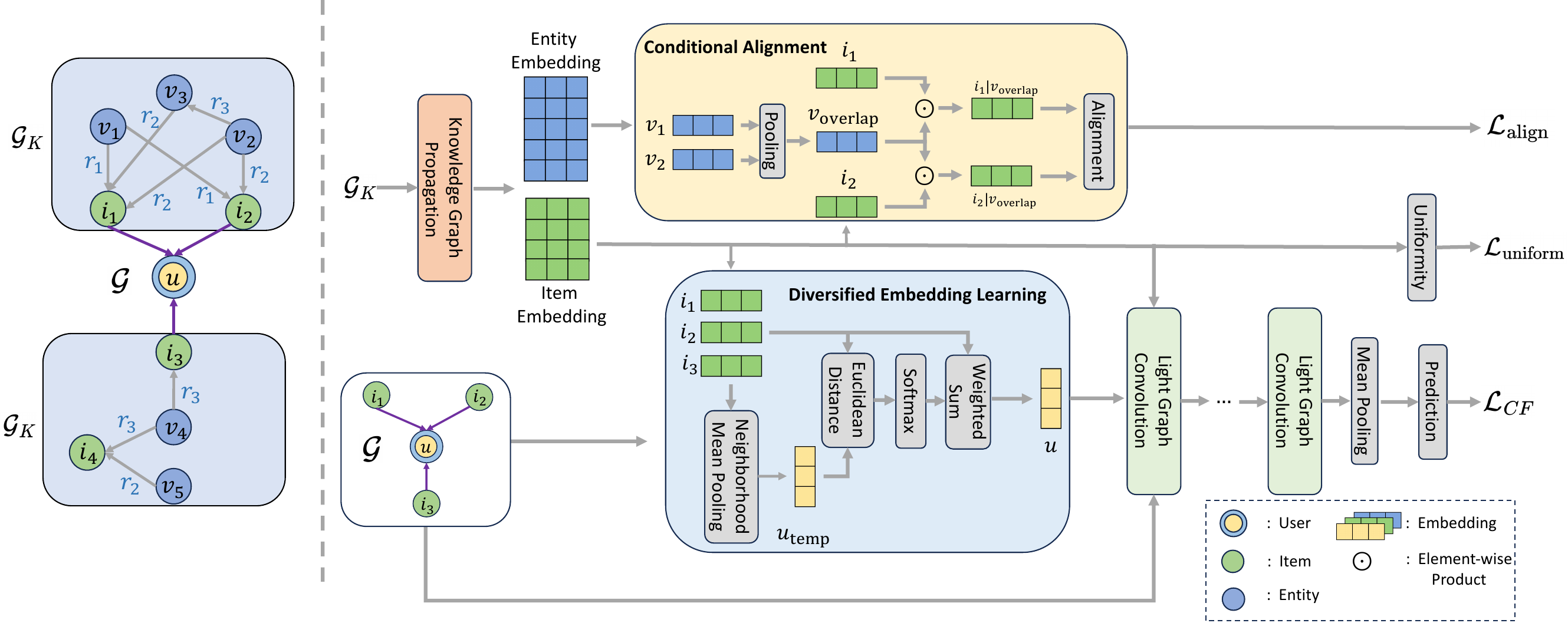}
    \caption{The framework of \modelname. 
    The leftmost figure is the collaborative knowledge graph.}

    \label{fig:mesh2}
\end{figure*}

The overall architecture of the proposed \modelname is displayed in Figure~\ref{fig:mesh2}, including knowledge graph propagation, diversified embedding learning and conditional alignment.

\subsection{PROBLEM FORMULATION}
We introduces the two data structures: user-item interaction graph, knowledge graph and the formulated problem.

\textbf{User-Item interactions graph.} 
We consider a collection of users denoted as $\mathcal{U}$, and a set of items represented by $\mathcal{I}$, where $m$ and $n$ correspond to the number of users and items, respectively. We establish a user-item bipartite graph $\mathcal{G} = \{\mathcal{V}, \mathcal{E}\}$, where $\mathcal{V} = \{\mathcal{U}, \mathcal{I}\}$ denotes the node set, and $\mathcal{E}$ represents the edge set. 
An edge $(u, i) \in \mathcal{E}$ signifies that the user $u$ purchased the item $i$ before.  

\textbf{Knowledge graph.}
Knowledge graph is defined as $\mathcal{G}_K = \{(h, r, t)\}$, where each triplet represents a relation $r \in \mathcal{R}$ from the head entity $h \in \mathcal{E}$ to the tail entity $t \in \mathcal{E}$.
Here, $\mathcal{E}$ and $\mathcal{R}$ represent the sets of entities and relations within KG, respectively.
Notably, the entity set $\mathcal{E}$ comprises both items ($\mathcal{I}$) and non-item entities ($\mathcal{E} \textbackslash \mathcal{I}$).

\textbf{Task Description.}
Given the $\mathcal{G}$ and $\mathcal{G}_K$, the goal of knowledge-aware recommendation is to recommend top $k$ items to each user.
Moreover, the diversified recommendation task in KG encourages more entities (or relations) covered by recommended items.

\subsection{Item entity-aware representation via Knowledge Graph propagation}
KG is a multi-relational graph comprising entities and relations, denoted by a set of triplets, i.e., $(h, r, t)$.
To facilitate understanding, we denote $N_i = \{(r, v)|(i, r, v) \in \mathcal{G}_K\}$ to represent the corresponding relations and entities connected to the item $i$.
A single entity may participate in multiple KG triplets, and it possesses the ability to adopt the linked entities as its attributes, thus revealing content similarity among items.
Take an example in Figure~\ref{fig:mesh1}, the star \textit{Chris Evans} involves in multiple KG triplets, i.e., (actor, \textit{The Avengers}), (actor, \textit{Before We Go}), and (director, \textit{Before We Go}). 
Then, we perform an aggregation technique to integrate the semantic information from $N_i$ to generate the representation of item $i$:
\begin{align}
    \textbf{e}_i^{(l)} = f(\{(\textbf{e}_i^{(l-1)}, \textbf{e}_r, \textbf{e}_v^{(l-1)})|(r,v) \in N_i\}),
\end{align}
where $\textbf{e}_i^{(l-1)}$ and $\textbf{e}_v^{(l-1)}$ represent the embedding of item $i$ and entity $v$ after $l-1$ layers aggregation respectively, and $\textbf{e}_r$ is the embedding of relation $r$.
$\textbf{e}_i^{(l)}$ is the aggregated representation obtained by $f(\cdot)$ which is the aggregation function that integrates information from ego-network $N_i$ of item $i$.
Regarding the function $f(\cdot)$, prior research efforts~\cite{KGAT} predominantly focus on incorporating the relation $r$ solely into the weight factors during propagation. Nevertheless, it is vital to acknowledge that the combination of connected relation and entity $(r, v)$ holds distinct contextual implications.
For instance, consider the star \textit{Chris Evans}, who is linked twice to the film \textit{Before We Go} in KG triplets, each time in a different role (director and actor). 
To address this contextual difference effectively, it becomes imperative to integrate relational contexts into the aggregator $f(\cdot)$. 
It not only facilitates discerning the diverse effects of relational nuances but also augments the overall diversity of relations within the model.
Therefore, we model the relation into the aggregation function $f(\cdot)$ following~\cite{KGIN}:
\begin{align}
    \textbf{e}_i^{(l)} = \frac{1}{|N_i|} \sum_{(r,v) \in N_i} \textbf{e}_r \odot \textbf{e}_v^{(l-1)},
    \label{encode}
\end{align}
where $\odot$ is the element-wise product. 
The relational message $\textbf{e}_r \odot \textbf{e}_v^{(l-1)}$ is propagated by modeling the relation $r$
as either a projection or rotation operator~\cite{RotatE}.
This design allows the relational message to effectively unveil distinct meanings inherent in the triplets.

\subsection{Diversified Embedding Learning}
After obtaining item embedding representation $\textbf{e}_i^{(l)}$ from $l$-th layer of knowledge graph, 
we leverage it to generate diversified user embedding.
We denote $\mathcal{N}_u = \{i|(u, i) \in \mathcal{E}\}$ to represent the $u$'s interacted items.
Then, we formulate the temporary user representation by applying mean pooling on the interacted items:
\begin{align}
    \textbf{e}^{(l)}_{u_{\text{temp}}} = \frac{1}{|\mathcal{N}_u|} \sum_{i \in \mathcal{N}_u} \textbf{e}_i^{(l)}.
    \label{intermediary}
\end{align}
In this way, the user representation is expressed by the representation of adopted items.
The motivation of conventional RecSys revolves around aligning the representations of users and their purchased items.
For example, in the left part of Figure~\ref{fig:mesh2}, the user $u$ interacted with $i_1$, $i_2$, and $i_3$.
A common approach involves driving the vector representation of user $u$ to approximate the vectors of items $i_1$ and $i_2$, as these two items display high similarity due to their overlapping connections on the KG.
However, it localizes the representation of $u$ in the surrounding area of $i_1$ and $i_2$ with their connected entities in KG (i.e., $v_1$, $v_2$, and $v_3$), which makes it hard to be exposed to the diverse items and entities around $i_3$ (i.e., $i_4$ and $v_5$).
Hence, we devise the user diversified embedding learning layer with the intent of liberating the user representation from localization constraints and fostering diversity.
The temporary user representation $\textbf{e}^{(l)}_{u_{\text{temp}}}$, where temp is short for temporary, is utilized to measure the dissimilarities with $\textbf{e}_i^{(l)}$ in the $l$-th layer:
\begin{align}
    & d_i = \| \textbf{e}^{(l)}_{u_{\text{temp}}} - \textbf{e}_i^{(l)} \|_2,\indent i \in \mathcal{N}_u, \\
    & a_i = \frac{\text{exp}(d_i)}{\sum_{j \in \mathcal{N}_u} \text{exp} (d_j)}.
\end{align}
Here, we utilize the Euclidean distance to compute dissimilarity, as it serves as a reliable measure to gauge the level of diversity. 
A higher distance value implies a greater potential for diversity in the resulting outcomes.
The abundance of similar items may confine the user representation, resulting in a resemblance to these item representations.
To counteract this, we mitigate the issue while preserving the intrinsic preference by generating diversity-aware embedding for the user through the following step:
\begin{align}
    \textbf{e}_u^{(l)} = \sum_{i \in \mathcal{N}_u} a_i \textbf{e}_i^{(l)}.
    \label{diversified embedding}
\end{align}
Consequently, the user representation is emancipated from the constraints of localization, which often result from an abundance of similar items.
For instance, $\textbf{e}_u$ will be compelled to distance itself from $\textbf{e}_i$ and $\textbf{e}_2$, while drawing nearer to $\textbf{e}_3$, thereby facilitating access to novel entities and items (e.g., $i_4$ and $v_5$).
Moreover, stacking multiple layers has been empirically demonstrated to effectively exploit high-order connectivity~\cite{KGAT, KGIN}.
Therefore, the $\textbf{e}_u^{(l)}$ can capture diverse information from $l$-hop neighbors of linked items, thereby enhancing exposure to diverse entities explicitly.
We further sum up the embeddings obtained at each layer to form the final user and item representation, respectively:
\begin{align}
    \textbf{e}_u = \textbf{e}_u^{(0)} + ... + \textbf{e}_u^{(L)},
    \indent \textbf{e}_i = \textbf{e}_i^{(0)} + ... + \textbf{e}_i^{(L)},
    \label{final_emb}
\end{align}
where $L$ is the number of knowledge graph propagation layers.

\subsection{Conditional Alignment and Uniformity}
To effectively encode KG in RecSys, we design a novel conditional alignment on KG entity embedding to preserve the semantic information between two similar items.
Consider item $i_1$ and item $i_2$, both sharing the entities $v_1$ and $v_2$.
Rather than a straightforward alignment~\cite{DirectAU,GraphAU}, the item representations $\textbf{e}_{i_1}$ and $\textbf{e}_{i_2}$ should be aligned based on the shared information of entity representations $\textbf{e}_{v_1}$ and $\textbf{e}_{v_2}$.
We define conditional embedding as follows:
\begin{align}
    \textbf{e}_{i_1|v_{overlap}} =  \textbf{e}_{i_1} \odot \textbf{e}_{v_{\text{overlap}}}, \indent
    \textbf{e}_{i_2|v} = \textbf{e}_{i_2} \odot \textbf{e}_{v_{\text{overlap}}},
\end{align}
where $O_{i_1,i_2}$ is the set of entities shared by item $i_1$ and item $i_2$ in KG and $\textbf{e}_{v_{\text{overlap}}} = \frac{1}{|O_{i_1,i_2}|} \sum_{v \in O_{i_1,i_2}}\textbf{e}_v$ is the average embedding of the overlapping entities between item $i_1$ and item $i_2$.
After modeling the overlapping entities as the projection or rotation on two items, we then perform the alignment loss:
\begin{align}
    \mathcal{L}_{\text{align}} = \mathbb{E} \| \textbf{e}_{i_1|v} - \textbf{e}_{i_2|v} \|^2,
    \label{align}
\end{align}
This design plays an essential role in pulling two similar items closer according to their connectivity on KG.
Furthermore, the uniformity loss~\cite{align} is also applied to scatter the embedding uniformly:
\begin{equation}
    \mathcal{L}_{\text{uniform}} = \log \mathbb{E}_{i, i^{'} \in \mathcal{I}} \textbf{e}^{-2\|\textbf{e}_i - \textbf{e}_{i^{'}}\|^2}.
    \label{uniform}
\end{equation}

\subsection{Model Prediction}
After performing $L$ propagation layers in KG, we further employ convolutional operation to capture collaborative signals from user-item interaction.
Due to the effectiveness and simple architecture of LightGCN~\cite{LightGCN}, where feature transformation and activation function are removed, we adopt its Light Graph Convolution layer (LGC) to encode the collaborative signals from user-item interactions:
\begin{equation}
    \begin{gathered}
        \mathbf{e}_{u}^{(k)} = \sum_{i \in \mathcal{N}_u} \frac{1}{\sqrt{\left | \mathcal{N}_u \right |} \sqrt{\left | \mathcal{N}_i \right |} } \mathbf{e}_{i}^{(k-1)}, \\
        \mathbf{e}_{i}^{(k)} = \sum_{u \in \mathcal{N}_i} \frac{1}{\sqrt{\left | \mathcal{N}_i \right |} \sqrt{\left | \mathcal{N}_u \right |} } \mathbf{e}_{u}^{(k-1)},
    \end{gathered}
    \label{lighgcn}
\end{equation}
where $\mathbf{e}_{u}^{(k)}$ and $\mathbf{e}_{i}^{(k)}$ are the embedding of user $u$ and item $i$ at the $k$-th LGC layer, respectively.
Meanwhile, $\mathbf{e}_{u}^{(0)}$ and $\mathbf{e}_{i}^{(0)}$ are the embedding obtained by Eq.~\ref{final_emb}.
Following $K$ layers of propagation, we integrate the embeddings acquired at each layer into the ultimate embeddings for both users and items:
\begin{align}
    \begin{aligned}
        \mathbf{e}_u = \frac{1}{K}\sum_{k=0}^{K} \mathbf{e}_{u}^{(k)},
        \mathbf{e}_i = \frac{1}{K}\sum_{k=0}^{K} \mathbf{e}_{i}^{(k)}.
    \end{aligned}
\end{align}
The embeddings produced by different LGC layers stem from distinct receptive fields, which also facilitates diversity from high-order neighbors.
We choose the BPR loss~\cite{rendle2012bpr} for the optimization:
\begin{align}
    \mathcal{L}_{CF} = - \sum_{(u, i, j) \in \mathcal{O}} \text{log} \sigma(\mathbf{e}_{u}^{\text{T}} \mathbf{e}_{i}- \mathbf{e}_{u}^{\text{T}} \mathbf{e}_{j}), 
    \label{CF}
\end{align}
where $\mathcal{O} = \{ (u, i, j)| (u, i) \in \mathcal{E}, (u, j) \in \mathcal{E}^{-} \}$. $\mathcal{E}^{-}$ is the set of unobserved interactions, and the $j$ is sampled from items that user $u$ has not interacted.
Finally, the overall loss function is defined as:
\begin{equation}
    \mathcal{L} = \mathcal{L}_{\text{CF}} + \lambda_1 \mathcal{L}_{\text{align}} + \lambda_2 \mathcal{L}_{\text{uniform}} + \lambda_3 {\left \| \Theta \right \|}_2^2,
    \label{loss}
\end{equation}
where $\lambda_1$ and $\lambda_2$ decide the weight of alignment and uniformity loss, respectively, and $\lambda_3$ denotes the parameter regularizing factor.

\section{Experiments}
This section aims to answer the 3 research questions (RQ).
\begin{itemize}[leftmargin=*]
    \item \textbf{RQ1:} How does \modelname perform compared to other state-of-the-art recommendation methods? 
    \item \textbf{RQ2:} Does every designed module play a role in \modelname?
    \item \textbf{RQ3:} What are the impacts of hyper-parameters: KG propagation layers $L$, alignment weight $\lambda_1$, and uniformity weight $\lambda_2$? 

\end{itemize}

\subsection{Experimental Settings}

\subsubsection{Datasets}
To evaluate the effectiveness of our \modelname against other baseline models, we perform experiments on three benchmark datasets: Amazon-book~\footnote{http://jmcauley.ucsd.edu/data/amazon/}, Last-fm~\footnote{https://grouplens.org/datasets/hetrec-2011}, and Movielens~\footnote{https://grouplens.org/datasets/movielens/}, which are widely used in previous works~\cite{KGAT, DSKReG, KACL,he2016ups}.
To ensure the quality of user-item interactions, the 10-core setting (preserve the users who have at least ten interactions) was adopted on all datasets as previous research.
We partitioned the historical interactions of each user into three sets: training (80\%), validation (10\%), and test (10\%). Table 1 presents detailed statistics for these datasets.

\begin{table}
  \caption{Statistics of the Datasets.}
  \label{table1}
  \begin{tabular}{l | c c c}
        \toprule
        \textbf{Dataset} & \textbf{Amazon-Book} & \textbf{Last.FM} & \textbf{Movielens} \\
        \hline
        \textbf{\#Users} & 70,679 & 1,872 & 37,385 \\

        \textbf{\#Items} & 24,915 & 3,846 & 6,182\\

        \textbf{\#Interactions} & 846,434 & 42,346 & 539,300 \\
        \hline
        \textbf{\#Entities} & 113,487 & 9,366 & 24,536 \\

        \textbf{\#Relations} & 39 & 60 & 20\\

        \textbf{\#Triplets}  & 2,557,746 & 15,518 & 237,155 \\

        \bottomrule
  \end{tabular}
\end{table}

\subsubsection{Baselines}
We compare \modelname with several representative methods as baselines:

\noindent \textbf{(1) General Recommender Systems}
\begin{itemize}[leftmargin=*]
    \item MF~\cite{MF} factorizes the user-item interaction matrix into user/item latent vectors via the BPR loss.
    \item LightGCN~\cite{LightGCN} is the state-of-the-art GNN recommender system. It is a simplified GCN~\cite{GCN} by removing the transformation function and non-linear activation.
    \item DirectAU~\cite{DirectAU} optimizes alignment and uniformity as the objective function to enhance collaborative filtering methods.
\end{itemize}

\noindent \textbf{(2) Knowledge Graph for Recommendation}
\begin{itemize}[leftmargin=*]
    \item KGAT~\cite{KGAT} utilizes an attention mechanism to discriminate the entity importance in KG during recursive propagation.
    \item KGIN~\cite{KGIN} uncovers the users' intents via the attentive combination of relations in KG and utilizes relational path-aware aggregation to refine user/item representations.
\end{itemize}

\noindent \textbf{(3) Diversified Recommender Systems}
\begin{itemize}[leftmargin=*]
    \item DGCN~\cite{DGCN} is the GNN-based method for diversified recommendation with rebalanced neighbor and adversarial learning.
    \item DGRec~\cite{DGRec} is the current state-of-the-art diversified RecSys model based on GNN. It designs the submodular function to select a diversified neighbors to enhance diversity.
\end{itemize}

\subsubsection{Evaluation Metrics}
In our experiments, we evaluate accuracy and diversity, respectively.

\indent \textit{Accuracy.}
We use Recall@k (R@k) and NDCG@k (N@k)~\cite{evaluation} with k ranges in \{20, 40\} to evaluate the performance of the top-k recommendation. 
The average metrics of all test users are reported.

\indent \textit{Diversity.}
Some studies~\cite{DivKG,survey} employ the \textit{ILAD} metric to quantify diversity based on dissimilarities among recommended items, yet its effectiveness heavily relies on the quality of learned embedding.
For assessing the diversity of recommendations on KG, we establish two metrics on KG: (1) Entity Coverage (EC) and (2) Relation Coverage (RC).
Entity/Relation Coverage refers to the number of entities/relations the recommended itemsets cover. 
We report top-20 and top-40 retrieval results to accord with accuracy.

\subsection{Performance Comparison (RQ1)}
\begin{table*}[]
\caption{Performance Comparison. The best and runner-ups are marked in bold and underlined separately.}
\label{performance comprision}
\centering
    \scalebox{1}{
\begin{tabular}{cl|cccc|cccc}
\hline
{\multirowcell{2}{Datasets}} & {\multirowcell{2}{Methods}} & \multicolumn{4}{c}{Accuracy}                                                 & \multicolumn{4}{c}{Diversity}                                                                            \\ \cline{3-6} \cline{7-10} 
 & {}                  & R@20   & R@40    &  N@20         & N@40      &  EC@20 & EC@40 & RC@20 & RC@40               \\ \hline
\multirow{6}{*}{{\multirowcell{2}{Amazon- \\ Book}}} & {MF} &  0.0999 & 0.1431 & 0.0524 & 0.0634 & 63.1383 & 97.2867 & \underline{20.2529} & 21.5524 \\ 
 & {LightGCN}  &  0.1090 & 0.1647 & 0.0558 & 0.0699 & 64.1135 & 95.1836 & 19.9618 & 21.8544  \\ 
 & {DirectAU}   &  0.1225 & 0.1716 & \underline{0.0650} & \underline{0.0776} & 48.0056 & 73.0553 & 18.0014 & 20.0823  \\ 
  & {KGAT}   &  \underline{0.1245} & \textbf{0.1846} & 0.0620 & 0.0773 & 52.4172 & 79.2991 & 18.4723 & 20.5078         \\ 
   & {KGIN}  & \textbf{0.1281} & \underline{0.1817} & \textbf{0.0673} & \textbf{0.0809} & 50.2582 & 75.2091 & 18.3580 & 20.2204            \\ 
 & {DGCN}    &  0.0805 & 0.1285 & 0.0407 & 0.0528 & 56.1095 & 84.3621 & 19.6098 & 21.5630        \\ 
    & {DGRec}    &  0.0920 & 0.1414 & 0.0458 & 0.0584 & \underline{65.2395} & \underline{97.5764} & 20.1816 & \underline{21.9992}        \\ 
    & {\modelname} & 0.1071 & 0.1621 & 0.0547 & 0.0687 & \textbf{67.8613} & \textbf{101.1123} & \textbf{20.5630} & \textbf{22.5242}   \\ 
 \bottomrule \bottomrule

\multirow{6}{*}{{\multirowcell{2}{Last.FM}}} & {MF} &  0.3292 & 0.4395 & 0.1802 & 0.2056 & \underline{68.2359} & \underline{122.5158} & 11.5179 & 14.3109 \\ 
 & {LightGCN}  &  \textbf{0.3685} & \textbf{0.4719} & \textbf{0.2016} & \textbf{0.2257} & 67.0864 & 117.2707 & 10.6190 & 13.4332 \\ 
 & {DirectAU}   &  0.3379 & 0.4349 & 0.1780 & 0.2005 & 60.4386 & 108.7641 & 8.8190 & 11.8342  \\ 
  & {KGAT}   &  0.3334 & 0.4373 & \underline{0.1818} & 0.2057 & 66.2712 & 118.7859 & 11.1397 & 14.0098         \\ 
   & {KGIN}  &  0.3415 & 0.4537 & 0.1635   & 0.1895 & 66.1098 & 117.9679 & \underline{11.8299} & \underline{14.5141}     \\
       & {DGCN}    &  0.2430 & 0.3474 & 0.1242 & 0.1482 & 59.8146 & 114.2739 & 10.7978 & 13.8157        \\ 
    & {DGRec}    &  0.2602 & 0.3618 & 0.1294 & 0.1529 & 60.5130 & 111.8712 & 11.1902 & 14.1957        \\ 
  & {\modelname}  &  \underline{0.3539} & \underline{0.4716} & 0.1785 & \underline{0.2059} & \textbf{87.2005} & \textbf{154.7543} & \textbf{13.9125} & \textbf{16.6293}    \\ 
 \bottomrule \bottomrule


   \multirow{6}{*}{{\multirowcell{2}{Movielens}}} & {MF} & 0.4065 & 0.5297 & 0.2187 & 0.2490 & \underline{454.3015} & \underline{805.9360} & 17.1941 & 17.3594 \\ 
  & {LightGCN}  & \underline{0.4157} & \underline{0.5462} & \underline{0.2236} & 0.2557 & 452.7703 & 803.3682 & 17.1809 & 17.3426 \\ 
 & {DirectAU}   & 0.4077 & 0.5256 &  0.2257 & 0.2547 & 441.1399 & 790.9293 & \textbf{17.2679} & \textbf{17.4271} \\ 
  & {KGAT}   &  0.4097 & 0.5340 & 0.2228 & 0.2533 & 444.9880 & 792.5169 & 17.2219 & 17.3757  \\ 
   & {KGIN} &  \textbf{0.4374} & \textbf{0.5669} & \textbf{0.2386} & \textbf{0.2705} & 447.2437 & 797.7831 & 17.1909 & 17.3531           \\ 
       & {DGCN}    &  0.3625 & 0.4941 & 0.1874 & 0.2198 & 444.5471 & 799.3097 & \underline{17.2358} & \underline{17.4053}   \\ 
    & {DGRec}    &  0.3676 & 0.4981 & 0.1917 & 0.2238 & 445.0291 & 798.6214 & 17.2310 & 17.3980   \\ 
 & {\modelname}  & 0.4132 & 0.5425 & 0.2241 & \underline{0.2559} & \textbf{458.8848} & \textbf{815.3123} & 17.1993 & 17.3710  \\ 
 \bottomrule \bottomrule
\end{tabular}
}
\end{table*}

\subsubsection{Parameter Setting}
We implement \modelname and all baselines on Pytorch with Adam~\cite{Adam} as the optimizer. 
We train each model until there is no performance improvement on the validation set within 10 epochs.
Following the setting of KGAT~\cite{KGAT}, we fix the embedding size to 64 for all models.
Grid search is performed to tune the hyper-parameters for each method.
The model parameters are initialized by Xavier initializer.
For KGAT, the pre-trained MF embeddings of users and items are used for initialization.

Specifically, the learning rate and weight decay are tuned in \{0.1, 0.05, 0.01, 0.005, 0.001\} and \{1e-4, 1e-5, 1e-6, 1e-7\} respectively. 
The number of KG layers and LightGCN layers are searched with the range of \{1, 2, 3, 4, 5, 6\}.
Finally, we tune the coefficients of alignment $\lambda_1$ and uniformity $\lambda_2$ among \{0.1, 0.2, ..., 1.0\} with increment of 0.1.

We report the overall performance of all baselines on the three datasets in Table \ref{performance comprision}. We have the following observations:
\begin{itemize}[leftmargin=*]

    \item \modelname outperforms all the other baselines on diversity metric: Entity Coverage (EC) and Relation Coverage (RC) except the RC on Movielens dataset. 
    However, all the performances on RC of Movielens dataset are almost the same, making models indistinguishable with each other regards RC.

    \item Despite achieving the highest Entity Coverage, \modelname shows comparable performance on Recall and NDCG metrics, particularly attaining the second-best results on Last.FM and exhibiting similarity to the second-best LightGCN on MovieLens dataset.
    This observation indicates that \modelname effectively enhances diversity while incurring a minimal cost to accuracy, thereby striking a well-balanced accuracy-diversity trade-off.

    \item DGRec is the state-of-the-art diversified RecSys that models diversity at the coarse-grained category information.
    Our model, by consistently outperforming DGRec across three datasets in all measures, especially on Last.FM, demonstrates the efficacy of modeling diversity at the fine-grained level of KG entities. It enables accurate recommendations and facilitates the provision of diverse and relevant information to a significant extent.

    \item In contrast to general accuracy-based methods (MF, LightGCN, and DirectAU), \modelname exhibits superior enhancements in diversity metrics while maintaining comparable performance in accuracy metrics. This observation serves as a testament to the model's effective capability in encoding KG.

\end{itemize}

\begin{figure}
     \centering
     \begin{subfigure}[b]{0.38\textwidth}
         \centering
         \includegraphics[width=\textwidth]{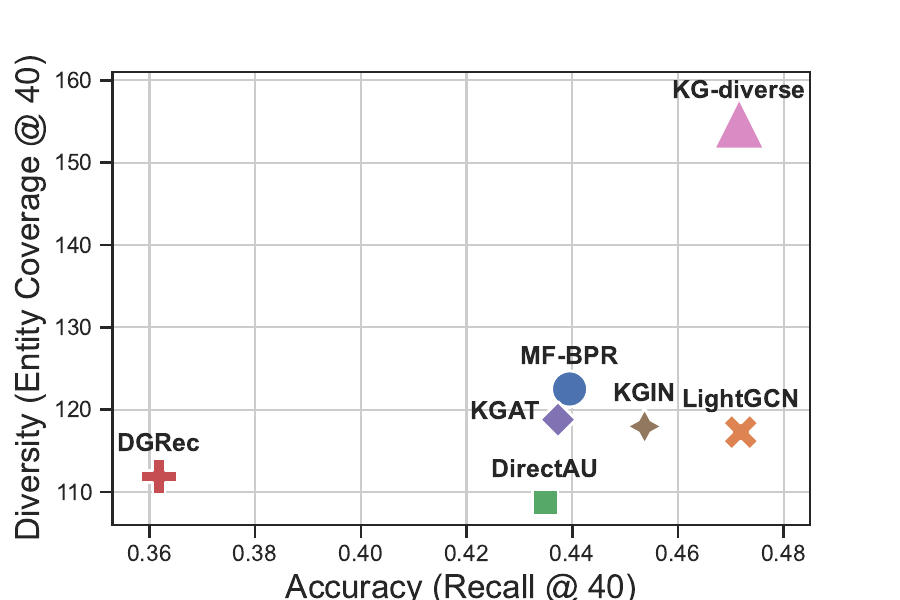}
         \label{music_accuracy_diversity}
     \end{subfigure}
    \caption{Accuracy-Diversity trade-off comparison. The upper-right model is better.}
        \label{trade-off}
\end{figure}

To be more specific, Figure~\ref{trade-off} illustrates the accuracy-diversity trade-off comparison of all models on Last-FM datasets, where Recall@40 and EC@40 measure accuracy and diversity, respectively.
It is evident from the plot that \modelname occupies the most favorable position in the upper-right quadrant, demonstrating that \modelname achieves the optimal accuracy-diversity trade-off.
Our model \modelname outperforms all other baselines in both accuracy and diversity metrics except LightGCN.
When compared to LightGCN, \modelname exhibits a substantial increase in diversity while making only a minor sacrifice in accuracy.

\subsection{Ablation Study (RQ2)}

We explore the effects of each module in \modelname on the Amazon-book dataset, including (1) KG propagation layer for item representation learning, 
(2) Relation encoding into item representation in Eq.~\ref{encode}, 
(3) Diversified Embedding Learning (DEL) module in generating user diversity-aware representations, and (4) Conditional Alignment and Uniformity (CAU) that regularizes KG embedding learning.
The results are shown in Table~\ref{Ablation study}:
\begin{itemize}[leftmargin=*]
    \item After removing the KG propagation layer, the knowledge within entities and relations will not be incorporated to enrich the item representations, and there is a single layer to generate user diversified representation without KG.
    Even though there are 2.7\% and 5.5 \% improvements on R@20 and N@20, respectively, the performance on EC@20 decreased by 7.1\%.
    It shows the effectiveness of learning entity-aware item representation via Eq.~\ref{encode}.

    \item Upon removing relation encoding, the item representation in $l$-th layer is obtained by averaging the embeddings of its connected entities in the previous layer. The results indicate significant decreases in all evaluation metrics, particularly a 5.9\% drop in RC@20.
    It indicates the crucial role of relations in enriching item representations.
    The presence of relations in the encoding process allows the model to capture and leverage the semantic connections between items and entities, contributing to more accurate and diverse recommendations.

    \item In the $l$-th KG propagation layer, the user diversity-aware embedding $\textbf{e}_u^{(l)}$ in Eq.~\ref{diversified embedding} is replaced by $\textbf{e}^{(l)}_{u_{\text{temp}}}$ in Eq.~\ref{intermediary} if DEL is removed.
    The declines in all four metrics: R@20 decreased by 4.3\%, N@20 decreased by 7.0\%, EC@20 decreased by 3.4\%, and RC@20 decreased by 2.1\%, underscores our model's advantage in enabling user representations to transcend localization constraints.

    \item Without conditional alignment and uniformity, there are substantial drops in R@20, N@20, and EC@20 metrics. This underscores the crucial role played by the regularization on KG embedding learning, as it serves as the foundational prerequisite for KG application.
    We find that the similarity and distinctions among items in the KG significantly influence the overall framework.
\end{itemize}

\begin{table}[t]
    \caption{Ablation Study.}
    \label{table4}
    \small
    \setlength{\tabcolsep}{1.2mm}{
    \scalebox{1}{
    \begin{tabular}{l|cccc}
         \hline
         Method & R@20 & N@20 & EC@20 & RC@20 \\
         \hline
        \modelname & \underline{0.1071} & \underline{0.0547} & \textbf{67.8613} &  \textbf{20.5630}\\
        \indent w/o KG propagation layer & \textbf{0.1101} & \textbf{0.0577} & 63.3891 & 20.1389 \\
        \indent w/o Relation encoding & 0.0902 & 0.0452 & \underline{66.4480} & 19.4177 \\
        \indent w/o DEL & 0.1027 & 0.0511 & 65.6006 & \underline{20.3592}\\
        \indent w/o CAU & 0.0929 & 0.0456 & 64.7545 & 20.2670\\
         \hline
    \end{tabular}
    }}
    \label{Ablation study}
\end{table}

\subsection{Parameter Sensitivity (RQ3)}
Additional exploratory experiments are conducted to comprehensively understand the \modelname structure and investigate the underlying reasons for its effectiveness.
We explore the effects of four essential hyper-parameters in \modelname on Last.FM dataset, including the number of KG propagation layers $L$, the coefficient $\lambda_1$ on alignment loss, and the weight decay $\lambda_2$ on uniformity loss.
\subsubsection{Number of KG propagation layers $L$}

We vary the number of KG propagation layers in the range of \{1, 2, 3, 4, 5, 6\} to analyze its influence in our model.
The two line plots reveal a clear trend: as the depth of KG propagation layers increases, the entity coverage steadily improves, reaching its peak at a depth of 5, and subsequently exhibits a decline. This trend signifies that the stacking of multiple KG layers aids item representations in capturing high-order connectivity among entities, thereby encoding a diverse range of entities within the item representations, as depicted in Equation~\ref{encode}.
In contrast, the accuracy metric (Recall) exhibits a consistent downward trend as the depth $L$ increases. It may be attributed to the over-smooth problem~\cite{over-smooth}, where the item embeddings become overly similar, negatively impacting the accuracy.

\begin{figure}
      \begin{center}
        \includegraphics[width=.22\textwidth]{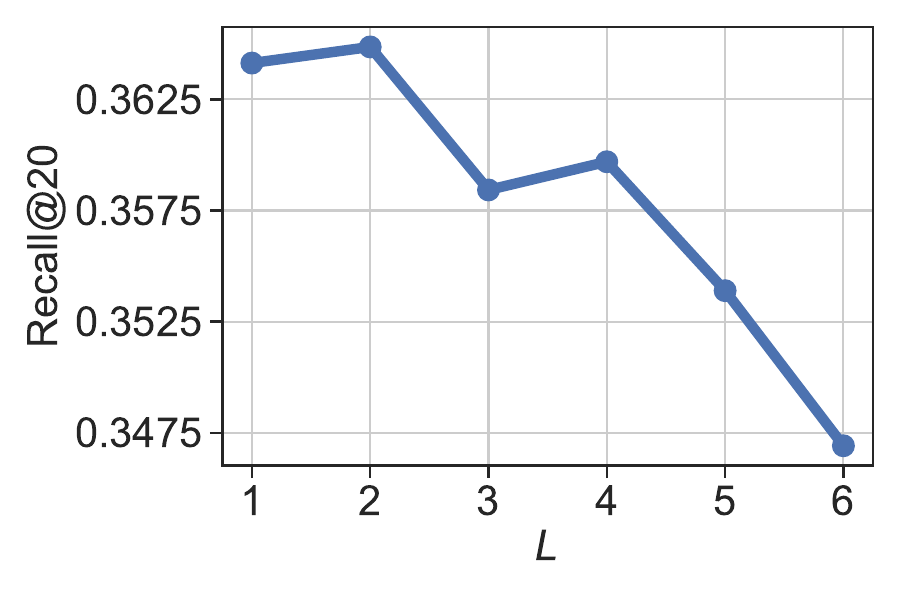}
        \includegraphics[width=.22\textwidth]{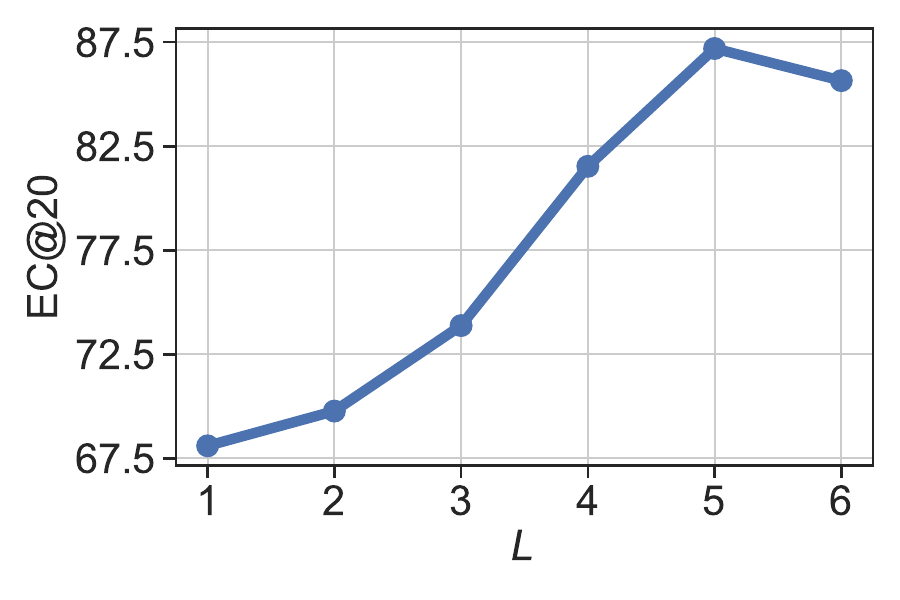}
        \includegraphics[width=.22\textwidth]{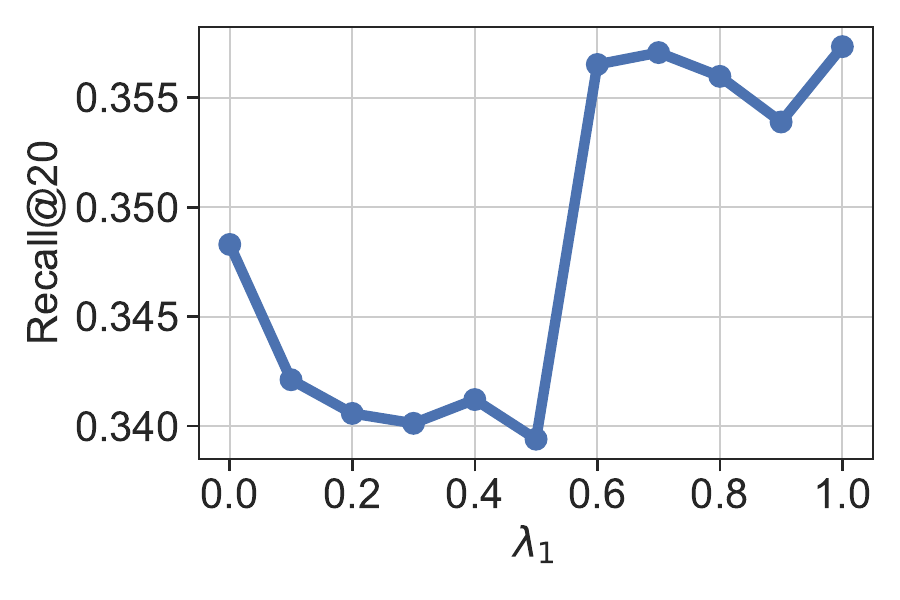}
        \includegraphics[width=.22\textwidth]{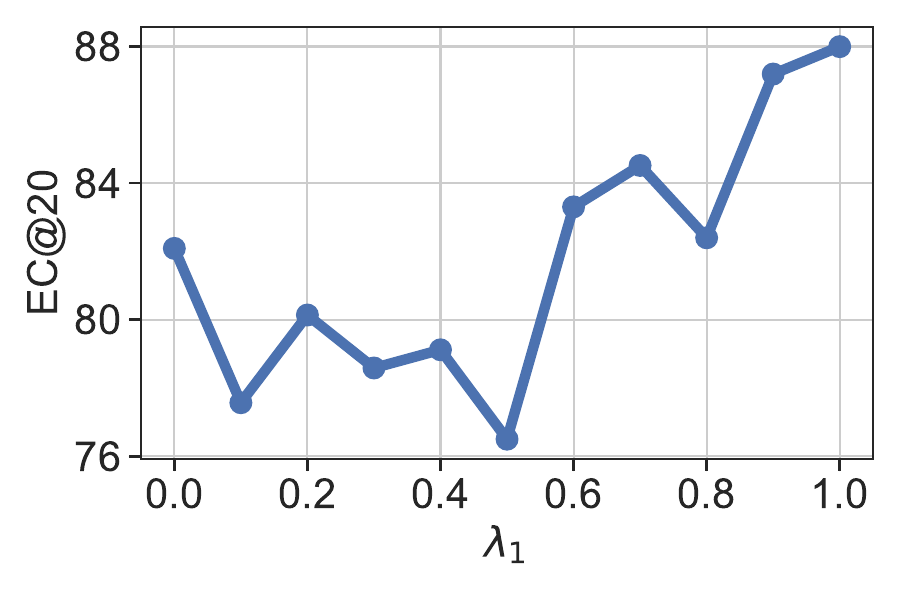}
        \includegraphics[width=.22\textwidth]{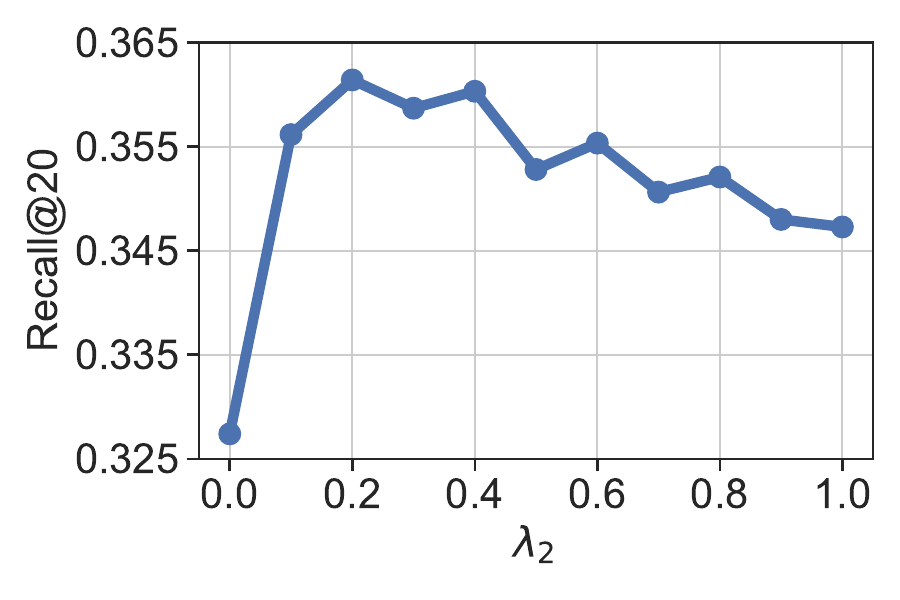}
        \includegraphics[width=.22\textwidth]{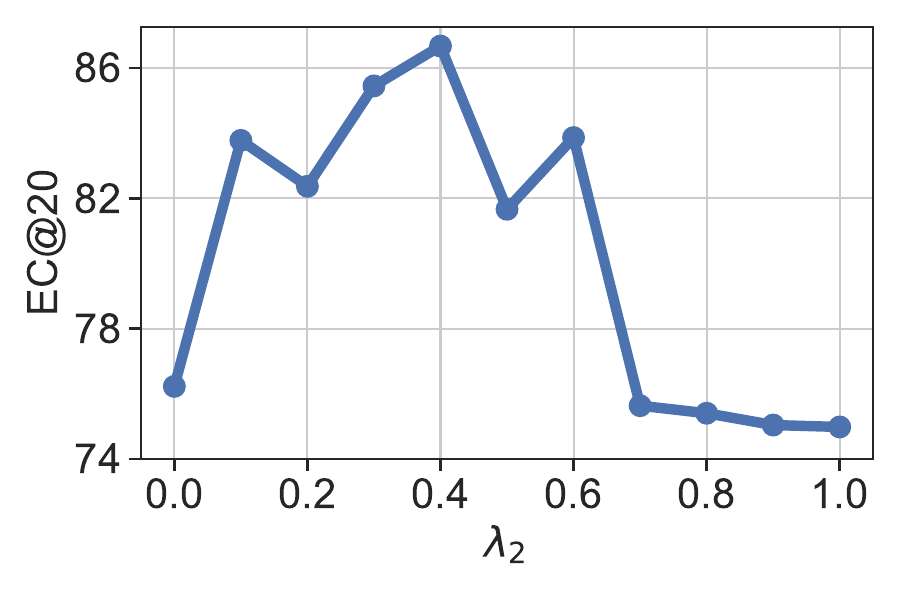}

      \end{center}
        \caption{Parameter sensitivity of $L$, $\lambda_1$, and $\lambda_2$}
        \label{Parameter sensitivity}
\end{figure}

\subsubsection{Coefficient $\lambda_1$ on alignment loss}.
The introduction of this coefficient aims to govern the weight of aligning item embeddings in the Knowledge Graph (KG). By adjusting the value of $\lambda_1$, \modelname becomes more attentive to aligning similar item representations, especially when two items share common entities in the KG.
As demonstrated in Figure~\ref{Parameter sensitivity}, the substantial enhancements observed in both Recall and Entity Coverage metrics serve as tangible evidence of the notable efficacy of conditional alignment in effectively encoding KG embeddings.
As similar items are successfully aligned, \modelname could easily retrieve the items that cover different and diverse entities.

\subsubsection{Weight decay $\lambda_2$ on uniformity}.
The incorporation of uniformity in \modelname serves the purpose of dispersing the item representations learned from the KG, preventing them from becoming identical and promoting diversity among the representations.
Our observations indicate that both accuracy and diversity exhibit an upward trend, achieving their optimal performance when the value of $\lambda_2$ is approximately 0.4. Nonetheless, as $\lambda_2$ increases further, the results for Recall and Entity Coverage witness a rapid decline.
This analysis leads us to the conclusion that a larger weight assigned to uniformity would promote more uniform embedding dispersal and potentially deteriorate KG alignment. However, a lighter emphasis on uniformity can still yield beneficial effects on the encoding of KG embeddings to some extent.

\subsection{Case Study}

\begin{figure}
     \centering
     \begin{subfigure}[b]{0.115\textwidth}
         \centering
         \includegraphics[width=\textwidth]{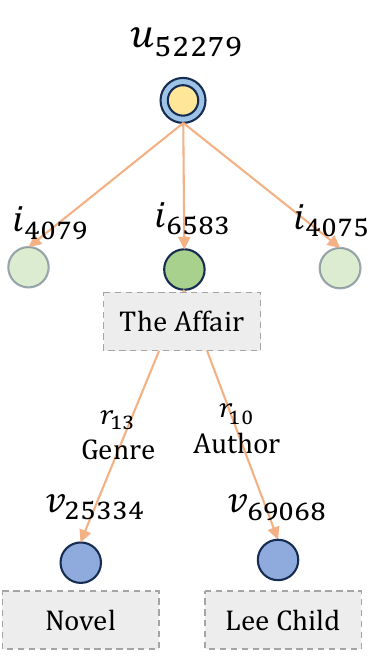}
         \caption{KGIN}
         \label{KGIN}
     \end{subfigure}
        \begin{subfigure}[b]{0.34\textwidth}
         \centering
         \includegraphics[width=\textwidth]{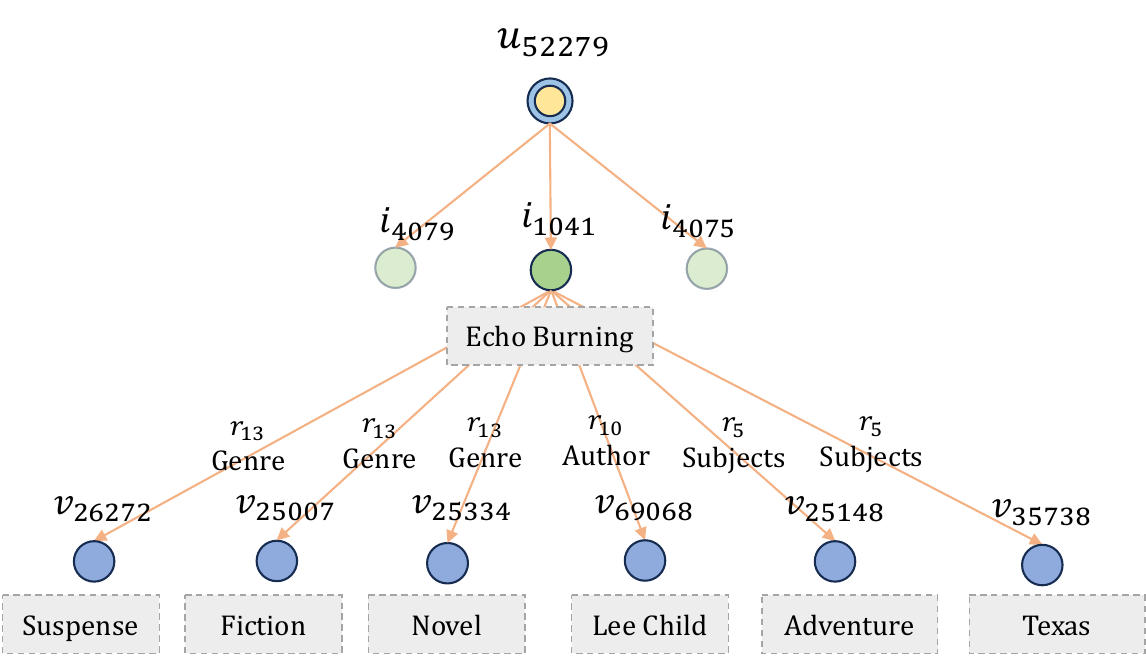}
         \caption{\modelname}
         \label{KG-diverse}
     \end{subfigure}
    \caption{Real Example from Amazon-Book.}
        \label{case study}
\end{figure}

In this case study, we randomly selecte a user, $u_{52279}$, from the Amazon-book dataset to compare the entity and relation coverages of two recommendation models: KGIN and \modelname. The visualization of their performance comparison is illustrated in Figure~\ref{case study}.
From both models, we retrieved the top-5 recommendations for $u_{52279}$, with {$i_{4079}, i_{4075}, i_{1153}, i_{3613}$} being common items recommended by both models. To simplify the analysis, we focus on two distinct recommended items: $i_{6583}$ from KGIN and $i_{1041}$ from \modelname.
We have the following observations:
\begin{itemize}[leftmargin=*]
    \item The item $i_{6583}$ with the name \textit{The Affair} covers only a few entities and relations, namely $N_{\text{The Affair}}$ = ((\textit{Genre}, \textit{Noval}), (\textit{Author}, \textit{Lee Child})). Remarkably, the item $i_{1041}$ recommended by our model, \modelname, also includes these two pairs of entities and relations. It exemplifies the strong relevance and effectiveness of \modelname in providing relevant recommendations.
    \item In addition to the strong relevance, the item $i_{1041}$ titled \textit{Echo Burning} covers a significantly broader range of entities and relations in the KG compared to $i_{6583}$. The entities and relations covered by $i_{1041}$ include {(\textit{Genre}, \textit{Suspense}), (\textit{Genre}, \textit{Fiction}), (\textit{Subjects}, \textit{Adventure}), (\textit{Subjects}, \textit{Texas})}. 
    It underscores the substantial capacity of \modelname in enhancing the diversity of recommendations. By recommending items like \textit{Echo Burning}, which encompass a wide variety of entities and relations, \modelname successfully introduces diversity and enriches the user experience by offering a more varied and engaging selection of recommendations.
\end{itemize}

\section{Related Works}
\subsection{Knowledge Graph for Recommendation}

CKE~\cite{CKE} model enriches the item representation by incorporating (1) structural embedding learned from items' structural knowledge via TransR, (2) textual and visual representations extracted by applying two different auto-encoders on textual and visual knowledge separately. 
DKN~\cite{DKN}, a news recommendation model, fuses semantic-level and knowledge-level information to news embedding by a multi-channel and word-entity-aligned knowledge-aware convolutional neural network (KCNN)~\cite{KCNN}. 
The user's vector is aggregated from the attention-weighted sum of news he/she clicked. 
KGIN~\cite{KGIN} models a finer-grained level of each intent by a combination of KG relations and designs a relational path-aware aggregation scheme that integrates information from high-order connectivities.
KGAT~\cite{KGAT} is designed to update embeddings by recursive propagation to capture high-order connectivities and compute the attention weights of a node’s neighbors which aggregate information from high-order relations during propagation.
KGCL~\cite{KGCL}  defines knowledge graph structure consistency score to identify items less sensitive to structure variation and tolerate noisy entities in KG. 
MetaKRec~\cite{wang2022metakrec} extracts KG triples into direct edges between items via prior knowledge and achieves improved performance with graph neural network.

While the mentioned works primarily concentrate on enhancing the accuracy of recommendations by leveraging KG, our work distinguishes itself by focusing on elevating the diversity of recommendations at a fine-grained level while ensuring accuracy.

\subsection{Diversified Recommendation}
Diversified recommendation technique is used to provide users with a diverse set of recommendations rather than a narrow set of highly similar items.
The concept of diversified recommendation was first proposed by Ziegler et al.~\cite{Ziegler}. They performed a greedy algorithm to retrieve the diversified top-K items. 
DUM~\cite{DUM} designs the submodular function to select items by a greedy method and balance the utility of items and diversity in the re-ranking procedure.
Determinantal point process (DPP)~\cite{DPP}, a post-processing method, retrieves the set of diverse items depending on the largest determinant of the item similarity matrix with MAP inference.
DGCN~\cite{DGCN} stands as the initial diversified recommendation approach based on GNNs. It adopts the inverse category frequency to select node neighbors for diverse aggregation and employs category-boosted negative sampling and adversarial learning to enhance item diversity within the embedding space.
DGRec~\cite{DGRec} is the current state-of-the-art model that produces varied recommendations through an enhanced embedding generation process. The proposed modules (i.e., submodular neighbor selection, layer attention, and loss reweighting) collectively aim to boost diversity without compromising accuracy.
DivKG~\cite{DivKG} learns knowledge graph embeddings via TransH~\cite{TransH} and designs a DPP kernel matrix to ensure the trade-off between accuracy and diversity.
EMDKG~\cite{EMDKG} encodes the diversity into item representations by an Item Diversity Learning module to reflect the semantic diversity from KG.

The previous works study diversity as the covered categories, which is measured at the coarse-grained level. This paper first investigates the fine-grained diversity with KG context information.

\section{Conclusion}
In this paper, we design a straightforward and effective framework named \modelname to enhance the diversity of recommendations with the rich information from KG. 
It first adopts the KG propagation method to incorporate semantic information into item representations.
Then, we propose a diversified embedding learning module to generate diversity-aware embedding for users.
Finally, we design Conditional Alignment and Uniformity to effectively encode the KG to preserve the similarity between two items with shared entities.
Furthermore, we introduce Entity Coverage (EC) and Relation Coverage (RC) to measure the diversity in KGs and evaluate the final performance. 
Extensive experiments on three public datasets demonstrate the effectiveness of \modelname in balancing the accuracy and diversity trade-off.

\section*{Acknowledge}
Hao Peng is supported by National Key R\&D Program of China through grant 2022YFB3104700, NSFC through grants 62322202, U21B2027, 62002007, 61972186 and 62266028, 
Beijing Natural Science Foundation through grant 4222030, S\&T Program of Hebei through grant 21340301D, Yunnan Provincial Major Science and Technology Special Plan Projects through grants 202302AD080003, 202202AD080003 and 202303AP140008, General Projects of  Basic Research in Yunnan Province through grants 202301AS070047, 202301AT070471, and the Fundamental Research Funds for the Central Universities. 
This work is supported in part by NSF under grant III-2106758. 

\bibliographystyle{ACM-Reference-Format}
\bibliography{reference}

\end{document}